\def\ignore#1{}

\def\nn#1 #2{#2. #1}				% Name with 1 initial
\def\nnn#1 #2 #3{#2. #3. #1}			% Name with 2 initials
\def\nnnn#1 #2 #3 #4{#2. #3. #4 #1}		% Name with 3 initials
\def\nnnnn#1 #2 #3 #4 #5{#2. #3. #4 #5. #1}	% Name with 4 initials

\def\pp{\noindent\parshape 2 0truecm 13.6truecm 1truecm 12.6truecm}
\def\rl#1;#2;#3;#4;#5 {\addtocounter{enumi}{1}\item[{[\arabic{enumi}]}]\par``#1", #2, {\it #3}, {\bf #4}, #5 \par}
\def\rlnonref#1;#2;#3;#4;#5 {\addtocounter{enumi}{1}\item[{ [\arabic{enumi}]}]\par``#1", #2, {\it #3}, {\bf #4}, #5 \par}
\def\pop#1;#2;#3; {\addtocounter{enumi}{1}\item[{[\roman{enumi}]}]\par``#1", {\it #2}, #3 \par}
\def\rn{\addtocounter{enumi}{1}\item[{[\arabic{enumi}]}]}

\def\rf#1;#2;#3;#4 {\par\pp#1, {\it #2}, {\bf #3}, #4. \par}
\def\rfprep#1;#2;#3 {{\par\frenchspacing\rn#1, #3 (#2).\par}}
\def\rk#1;#2;#3;#4;#5 {\par``#1", #2, {\it #3}, {\bf #4}, #5 \par}
\def\ro{}

\def\beq#1{\begin{equation}\label{#1}}
\def\eeq{\end{equation}}
\def\beqa#1{\begin{eqnarray}\label{#1}}
\def\eeqa{\end{eqnarray}}
\def\eq#1{equation~(\ref{#1})}
\def\Eq#1{Equation~(\ref{#1})}
\def\eqn#1{~(\ref{#1})}

\def\spose#1{\hbox to 0pt{#1\hss}}
\def\simlt{\mathrel{\spose{\lower 3pt\hbox{$\mathchar"218$}} \raise 2.0pt\hbox{$\mathchar"13C$}}}
\def\simgt{\mathrel{\spose{\lower 3pt\hbox{$\mathchar"218$}} \raise 2.0pt\hbox{$\mathchar"13E$}}}
\def\simpropto{\mathrel{\spose{\lower 3pt\hbox{$\mathchar"218$}} \raise 2.0pt\hbox{$\propto$}}}

\def\bt{\begin{tabbing}}
\def\et{\end{tabbing}}
\def\beq#1{\begin{equation}\label{#1}}
\def\eeq{\end{equation}}

\def\Sec#1{Section~\ref{#1}}

\def\bfig{\begin{figure}[h] \centerline{\hbox{}}\vfill}
\def\efig{\end{figure}\vfill\newpage}

\def\fig#1{Figure~\ref{#1}}

\def\fig#1{Figure~\ref{#1}}
\def\Fig#1{Figure~\ref{#1}}

\def\Cl{C_\l}

\def\expec#1{\langle#1\rangle}

\def\l{\ell}
\def\etal{{\frenchspacing\it et al.}}
\def\ie  {{\frenchspacing\it i.e.}}
\def\eg  {{\frenchspacing\it e.g.}}
\def\etc {{\frenchspacing\it etc.}}

\def\uK{\mu{\rm K}}

\def\abar{\bar{a}}
\def\almbar{\abar_{\ell m}}
\def\alm{a_{\ell m}}
\def\Ylm{Y_{\ell m}}

\def\figsize1{4.0cm}
\def\figsize{9.0cm}

\def\x{{\bf x}}

\def\W{\rm \bf W}

\def\a{{\bf a}}
\def\n{{\bf n}}
\def\r{{\bf r}}
\def\ahat{\widehat{\a}}
\def\rhat{\widehat{\r}}
\def\almhat{\widehat{a}_{\l m}}

\def\zero{{\bf 0}}

\def\Y{{\bf Y}}
\def\C{{\bf C}}
\def\I{{\bf I}}
\def\N{{\bf N}}
\def\S{{\bf S}}
\def\SS{{\bf \Sigma}}

\def\SSS{\SS_S}

\documentclass[prd,twocolumn,amsmath,nofootinbib]{revtex4} % For astro-ph
\usepackage{url}

\begin{document}
\input{epsf.sty}

\title{CMB multipole measurements in the presence of foregrounds}

\author{Ang\'elica de Oliveira-Costa \& Max Tegmark}
\address{MIT Kavli Institute for Astrophysics and Space Research, Cambridge, MA 02139; angelica@mit.edu}
\date{\today. to be submitted to Phys. Rev. D.}


\begin{abstract}
Most analysis of Cosmic Microwave Background spherical harmonic coefficients $a_{\l m}$
has focused on estimating the power spectrum $C_\l=\expec{|a_{\l m}|^2}$ rather than 
the coefficients themselves. We present a minimum-variance method for measuring 
$a_{\l m}$ given anisotropic noise, incomplete sky coverage and foreground contamination,
and apply it to the WMAP data. Our method is shown to constitute lossless data compression 
in the sense that the widely used quadratic estimators of the power spectrum $C_\l$ can 
be computed directly from our $a_{\l m}$-estimators.
As the Galactic cut is increased, the error bars $\Delta a_{\l m}$ on low multipoles 
go from being dominated by foregrounds to being dominated by leakage from other 
multipoles, with the intervening minimum defining the optimal cut. 
Applying our method to 
the WMAP quadrupole and octopole as an illustration, we investigate the robustness of the previously reported ``axis of evil'' 
alignment to Galactic cut and foreground contamination.
\end{abstract}

\maketitle

\section{INTRODUCTION}

Cosmology has been revolutionized by the advent of precision maps of the Cosmic Microwave 
Background (CMB), allowing accurate tests of the Cosmological Standard Model (CSM) and 
measurements of its free parameters (\eg, \cite{spergel03,sdsspars,sdsslyaf,2dfpars,b2kpars}).
Although the analysis of CMB maps has traditionally focused on the power spectrum, various 
surprises discovered in the data have triggered a growing interest in extracting additional 
information pertaining to possible non-Gaussianity. Specifically, the all-sky CMB 
temperature fluctuation map $\delta T(\rhat)$ is customarily expanded in spherical harmonics:
\beq{almDefEq}
\delta T(\rhat) = \sum_{\l m}\alm\Ylm(\rhat).
\eeq
These multipole coefficients $\alm$ are treated as stochastic variables, varying randomly 
when extracted from CMB maps in widely separated Hubble volumes. According to the CSM, 
the CMB fluctuations are for all practical purposes Gaussian and statistically isotropic,
which implies that the $\alm$-coefficients are independent Gaussian random variables with 
zero mean, \ie, that the CMB contains no cosmological information whatsoever except for 
its power spectrum $\Cl=\expec{|\alm|^2}$.

Although some inflation models predict potentially observable departures from Gaussianity on 
small angular scales ($\l\simgt 10^2$), no such non-Gaussianity has yet been found (see 
\cite{komatsu03,nongaussianity} and references therein).
However, a number potential non-Gaussian anomalies have recently been discovered on large angular scales.
The surprisingly low CMB quadrupole $C_2$ has intrigued the cosmology community ever since it was first 
observed by COBE/DMR \cite{smoot92}, and was taken more seriously once the precision measurements of 
WMAP \cite{bennett03b} showed that it could not be complete blamed on Galactic foreground contamination. 
Although the low quadrupole does not involve non-Gaussianity and its statistical significance is debatable 
	% \cite{spergel03,verde03,enrique03,efstathiou03b,seljak?}, 
	  \cite{spergel03,enrique03,efstathiou03b,seljak04}, 
it is a fly in the ointment of the CSM, and closer investigation of this has revealed numerous hints of 
non-Gaussianity on large angular scales.

The first reported hints of such non-Gaussianity involved the quadrupole and octopole. They were both found 
to be rather planar, \ie, with most of their hot and cold spots centered in a single plane in the sky, with 
their two preferred planes surprisingly closely aligned \cite{tegmark03,smalluniverse}.
In the CSM context, the reported alignment required a 1-in-66 fluke and the octopole planarity an independent 
1-in-20 coincidence \cite{smalluniverse}. These features were confirmed by other groups and examined with other 
techniques, finding puzzling alignments all the way up to $\l=5$ that may require as much as a 1-in-1000 fluke
\cite{eriksen05a,copi05a,dominik,eriksen05b,smalluniverse,land1,weeks,land2,land3,bielewicz05,land4,copi05b,land5}.
The preferred axis, dubbed the ``axis of evil'' by \cite{land2}, points towards Virgo and is intriguingly 
close to the ecliptic poles \cite{copi05a,dominik,copi05b}. Additional hints of low-$\l$ non-Gaussianity have 
also turned up involving, \eg, asymmetries 
\cite{park04,eriksen05a,copi05a,eriksen05b,bielewicz04,naselsky04,land1,neil04,prunet,eriksen04,weeks,land2,land3,
gluck,bielewicz05,land4,copi05b,chen,land5,freeman,amir03,dore04,portsmouth,silk,eriksen04c,stannard,seto,bernui,bernui06}.

\begin{figure*}[t]
 \centerline{\epsfxsize=18cm\epsffile{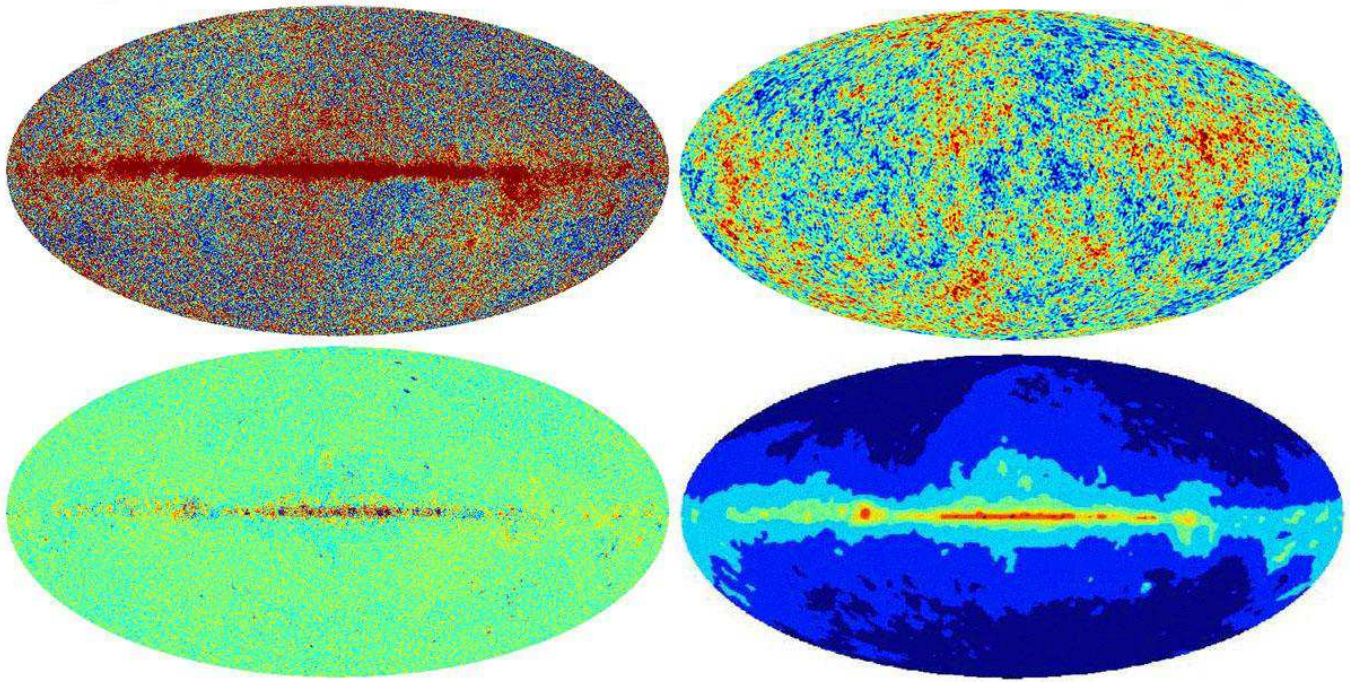}} 
\caption[1]{\label{mapFig}\footnotesize%
	Foreground contamination is clearly visible in the WMAP V-band map (top left).
	After taking a simulated WMAP CMB map (top right), adding simulated noise and 
	foregrounds, applying the TOH cleaning method \cite{tegmark03} and subtracting 
	the CMB back out, foregrounds remain clearly visible in the residual map (bottom 
	left). We apply our multipole measurement technique after masking out each of 
	seven regions of decreasing cleanliness (bottom right) to optimize the tradeoff 
	between residual foreground contamination and limited sky coverage (which causes 
	leakage from unwanted multipoles coupling to our estimator).
	}
\end{figure*}

These purely observational results have triggered numerous papers on potential physical explanations. These 
include effects of Galactic foreground emission \cite{eriksen04b,seljak04,naselsky,bernardi,medeiros}
and local structures \cite{raul,dominik,calvao,tomita,vale,asantha,alex06,scott06,silk06}
as well as theoretical explanations going beyond the CSM and involving compact cosmic topology 
\cite{smalluniverse,desitter,mota04,neil03,gomero03,roukema,janna,amir05,mota05a,mota05b,mota05c,gomero05,cresswell06,reboucas06},
modified inflation 
\cite{feng03,huang03,bi03,patricia03a,patricia03b,hogan03,kaloper,liguori,hogan04,bunity04,hunt,andrei,piao,buniy05}, 
or other new physics 
\cite{kostelecky,multamaki,string,darkenergyvector,gordon04,moffat,battye05,jaffe05,knox,gordon05b,gordon05a,sean,jain,koi05,sub06,cayon06,moss06,greek06}.

A substantial fraction of this work has involved the $\alm$-coefficients from \eq{almDefEq} with $\l\le 5$.
Given this interest in the low-$\l$ multipoles, it is timely and worthwhile to measure them as accurately 
as possible, with quantified error bars, further improving on approaches such as \cite{tegmark03,bielewicz04}.
This is the purpose of the present paper. More accurate $\alm$-measurements can either build confidence in 
the CSM or teach us about interesting new physics.

Because of Galactic foreground contamination and anisotropic detector noise noise, 
the best way to measure the multipole coefficients $\alm$ is not to simply apply
the relation $\alm=\int\Ylm(\rhat)^*\delta T(\rhat)d\Omega$ to a CMB map (see \fig{mapFig}).
Rather, the customary approach is to discard the most contaminated part of the sky and use a more
elaborate linear weighting on the remainder of the map. In \Sec{MethodSec}, we
study how to optimize this procedure. We then apply our method to the latest WMAP data in 
\Sec{ApplicationSec}, investigate the implications for the quadrupole-octopole alignment
in \Sec{AlignmentSec} and summarize our conclusions in \Sec{ConcSec}.

\section{Basic method}\label{MethodSec}

\subsection{The problem}

Given a CMB map with pixels $i=1,...,N$, let $x_i$ denote the observed temperature 
fluctuation $\delta T$ in the pixel whose sky direction corresponds to the unit vector 
$\rhat_i$. For the case of WMAP, there are $N=12\times 512^2 = 3$,$145$,$728$ pixels
distributed according to the HEALPix scheme\footnote{
    The HEALPix package is available from \protect\url{http://www.eso.org/science/healpix/}.
    }
\cite{healpix1,healpix2}, but we will only use those outside of some appropriate Galaxy 
cut below, and at a lower resolution. Suppose we wish to measure a particular set of 
multipole coefficients $a_{\l_j m_j}$, $j=1,...,M$, labeled simply $a_j$ for 
brevity\footnote{For convenience, since CMB maps are real rather than complex-valued, we work with
real-value spherical harmonics $\Ylm$ throughout this paper.
These are obtained from the standard spherical harmonics by replacing $e^{im\phi}$
by $\sqrt 2\sin m\phi$, $1$, $\sqrt 2\cos m\phi$ for $m<0$, $m=0$, $m>0$, respectively.
As detailed in Appendix A of \cite{smalluniverse}, the corresponding real-valued coefficients $\alm$ are related to
the traditional complex-valued coefficients $\almbar$ by a simple unitary transformation:
for $m<0$, $m=0$ and $m>0$,
$\almbar$ equals $\sqrt{2}\>{\rm Im}\>\alm$, $a_{\l 0}$ and $\sqrt{2}\>{\rm Re}\>\alm$, respectively.
This means that for $m<0$, $m=0$ and $m>0$,
the traditional complex coefficients $\almbar$ equal
$(-1)^m(\alm-ia_{\l,-m})/\sqrt{2}$, $\alm$ and $(\alm+ia_{\l,-m})/\sqrt{2}$, respectively.
}.
Grouping the pixels into an $N$-dimensional vector $\x$ and the multipole coefficients 
into an $M$-dimensional vector $\a$, we can rewrite \eq{almDefEq} as
\beq{ProblemEq}
     \x = \Y\a + \n,
\eeq
where the $N\times M$ spherical harmonic matrix $\Y$ is defined by
$\Y_{ij}=Y_{\l_j m_j}(\rhat_i)$ and the ``noise'' vector $\n$ contains 
all contributions to the sky map except from the desired multipoles.
In other words, $\n$ contains not only detector noise, but also 
genuine sky signal contributed by other multipole coefficients that are not 
included in the $\a$-vector. We make the usual assumption that the noise 
has zero mean ($\expec{\n}=\zero$) and define its covariance matrix 
$\C\equiv\expec{\n\n^t}$. If the detector noise covariance matrix is 
$\N$, then we can write this in the usual form
\beq{Ceq}
    \C=\N+\S,\quad \S_{ij}=\sum_\l {2\l+1\over 4\pi} P_\l(\rhat_i\cdot\rhat_j)C_\l,
\eeq
where $\S$ is the CMB contribution, $P_\l$ is a Legendre polynomial and and 
the sum runs over those multipoles that are not included in our $\a$-vector.
We ignore the issue of foreground contamination for now, but will cover this
important issue in detail below in \Sec{ApplicationSec}.

We wish to find an estimator $\ahat$ of the true multipole vector $\a$ that
is unbiased ($\expec{\ahat}=\a$) and whose elements have as low variance as 
possible.

\subsection{The solution}

For our low-$\l$ applications, the linear system given by \eq{Ceq} is normally 
greatly overdetermined with $N\gg M$, \ie, with many more pixels than desired 
multipole coefficients. Mathematically equivalent linear problems frequently 
occur in the CMB literature in the contexts of mapmaking and foreground removal, 
and the optimal solution is well-known to be \cite{mapmaking}
\beq{ahatEq}
    \ahat = \W\x,\quad \W\equiv [\Y^t\C^{-1}\Y]^{-1}\Y^t\C^{-1},
\eeq
with covariance matrix 
\beq{CovEq}
    \SS\equiv\expec{\ahat\ahat^t}-\expec{\ahat}\expec{\ahat}^t = [\Y^t\C^{-1}\Y]^{-1}.	 
\eeq
To make this method useful in practice, we need to augment it with a prescription 
for how to optimize the Galaxy cut when faced with foreground contamination ---  
we will do this in \Sec{ApplicationSec}. Before that, let us now provide some 
intuition for how the method works.

First note that since the spherical harmonics are orthogonal over the full 
sphere, complete sky coverage with uniform detector noise per pixel will make 
the covariance matrix $\SS$ diagonal, reducing our method to simply expanding 
the map in spherical harmonics the usual way. The measurement of a given harmonic 
$\alm$ is then the exact sky signal plus a detector noise contribution.

If part of the sky is missing due to foreground masking or lack of observation, 
or if the pixel weighting varies because of nonuniform detector noise, then 
the spherical harmonics are no longer orthogonal, which is reflected in $\SS$ 
from \eq{CovEq} having off-diagonal elements. However, the matrix algebra corrects 
for this coupling between the multipole coefficients. \Eq{ahatEq} implies that 
$\W\Y=\I$ so that the measurement error $\ahat-\a=\W(\Y\a+n)-\a =\W\n$, independent 
of $\a$. For example, if $\a$ contains the 16 multipole coefficients with $\l\le 3$,
then the measured quadrupole coefficient $\widehat{a}_{20}$ will equal the true 
full-sky value $a_{20}$ plus a noise contribution. This ``noise'' will include 
detector noise and leakage from sky multipoles with $\l>3$, but there will be 
no contribution whatsoever from the monopole, dipole or octopole, nor from quadrupole 
coefficients $a_{2m}$ with $m\ne 0$. The method will generically be able to solve for these 16 
unknowns as long as there are at least 16 pixels, but the error bars will clearly 
grow as the Galaxy cut is increased, since this makes it more difficult to disentangle the 
different multipoles. Numerically, we indeed find that the matrix $[\Y^t\C^{-1}\Y]$ remains nonsingular for all the Galaxy cuts
we consider, so no regularization techniques are needed. However, we will see that the error bars grow sharply as the cut 
increases and the matrix becomes less well-conditioned, particularly
because multipoles with large $|m|$ live predominantly in the masked-out regions near the Galactic plane.

\subsection{Relation to other methods}

The multiple estimation paper most closely related to this one is that of \cite{bielewicz04}.
They consider two alternative linear techniques that, cast in the notation of the present paper, 
simply use different $\W$-matrices than the one given by \eq{ahatEq}, 
perform a careful and detailed study of their statistical properties, and apply them to the WMAP data.
The first technique they explore is Wiener filtering, defined by 
$\W\equiv \Y^t\S^{-1}[\Y^t\C^{-1}\Y]^{-1}$.
Using numerical simulations, they confirm that this causes a 
systematic suppression of power with increasing $\l$ (decreasing signal-to-noise) as expected.
They then focus on the power equalization (PE) filter defined by a
$\W$-matrix based on Cholesky decomposing the spherical harmonic coupling matrix $\Y^t\Y$. 
The PE filter has the attractive property of eliminating all contributions
to a given multipole estimator from lower multipoles, but it is neither unbiased (in the sense that $\W\Y\ne\I$) nor minimum-variance.
For the special case where we include all multipoles up to some $\l$ in the $\a$-vector, our estimator 
for the last $\alm$-coefficient will also be independent of the lower multipoles, merely with a lower variance 
than the PE estimator (since the estimator defined by \eq{ahatEq} by construction gives the smallest variance of any estimator with this property).

\subsection{Relation to power spectrum estimation}

The most common technique in the CMB community for measuring CMB power spectra uses 
so-called quadratic estimators \cite{cl,bjk}, largely because they have been shown 
to be information-theoretically optimal, giving the smallest possible error bars.
Equations\eqn{ahatEq} and\eqn{CovEq} imply that
\beq{LosslessEq}
\ahat^t\SS^{-2}\ahat = \x^t\C^{-1}\Y\Y^t\C^{-1}\x.
\eeq
When $\Y$ contains all spherical harmonics for a given 
$\l$, the right hand side is precisely the quantity that a quadratic estimator of 
$\Cl$ extracts from the map $\x$. \Eq{LosslessEq} therefore shows that we can calculate 
optimal estimators of the power spectrum from our measured multipoles $\ahat$ without 
recourse to the CMB map. In other words, our method can be viewed as a form of lossless 
data compression, with all information from the map $\x$ about the power spectrum 
coefficient $\Cl$ retained in the measured coefficients $\ahat$.

\section{Application to WMAP}\label{ApplicationSec}

Let us now apply our method to the WMAP data \cite{bennett03b}.
After briefly describing the data and foreground masks used, we
present results for a variety of Galactic cuts. To quantify the 
foreground contribution and select the best cut, we then create 
and analyze simulated CMB and foreground maps. This also allows 
us to optimize other practical aspects of our method such as the 
pixel size.

\subsection{Data \& foreground masks used}

Our analysis is based on the 1 year WMAP data described in \cite{bennett03b}
with the maps from the five observing frequencies combined into the single foreground-cleaned map
TOH map described in \cite{tegmark03} and downloadable at \url{http://www.lambda.gsfc.nasa.gov}.
The TOH map is an all-sky CMB map with the same resolution as the highest-frequency WMAP channel 
(about 12.6' FWHM), HEALPix-pixelized at resolution level {$nside$}=512.
The TOH foreground cleaning algorithm assumes that the 
CMB has a blackbody spectrum, but is otherwise completely blind, making no assumptions 
about the CMB power spectrum, the foregrounds, WMAP detector noise or external templates. 

We explore the same series of increasingly conservative foreground masks used in \cite{tegmark03}
(\fig{mapFig}, bottom right). Their construction is described in detail in \cite{tegmark03}, 
and involves the following key steps. One first converts the maps at the five WMAP frequency 
bands K, Ka, Q, V and W to a common angular resolution and forms four difference maps W-V, 
V-Q, Q-Ka and Ka-K, thereby obtaining maps guaranteed to be free of CMB signal that pick up 
any signals with a non-CMB spectrum. One then creates a combined ``junk map'' (foreground map) 
by taking the largest absolute value of these four maps at each pixel. Finally, with appropriate 
smoothing, one creates sky regions based on contour plots of this map. We use cuts that are 
roughly equispaced on a logarithmic scale, corresponding to thresholds of 30000, 10000, 3000, 
1000, 300 and 100$\mu$K (\fig{mapFig}, right bottom). We label our cuts as masks 0, 1, ..., 8, 
where mask 8 uses pixels with less than 100 $\mu$K in the junk map and mask 0 uses the entire 
sky. Masks 1 \& 2 merely cut out small blobs in the Galactic plane as described in \cite{tegmark03}.
For masks 0,...,8, the total sky percentages removed are 0\%, 0.03\%, 0.08\%, 0.25\%, 0.64\%, 
2.0\%, 6.6\%, 24\% and 64\%, respectively.

\begin{figure} 
\centerline{\epsfxsize=\figsize\epsffile{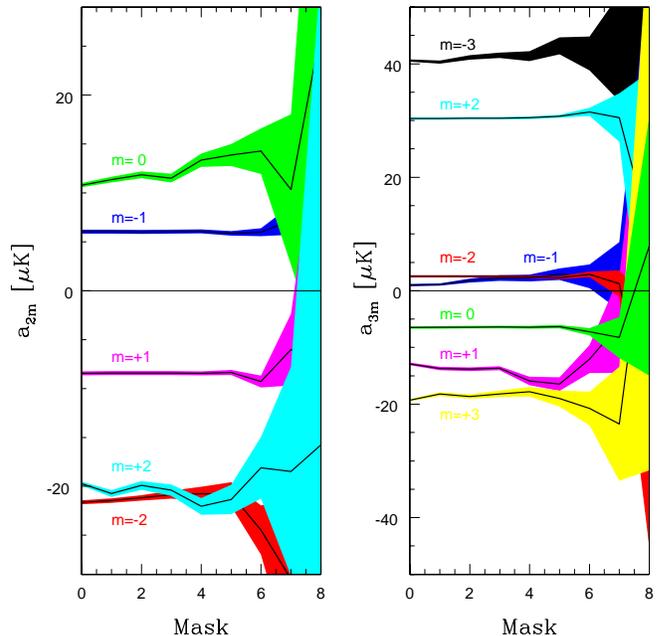}}
\caption[1]{\label{tornadoFig}\footnotesize%
	The left and right panels show how the $\protect\alm$-coefficients 
	(phases) of the quadrupole (left) and the octopole (right) extracted 
	from the TOH map change as more sky is masked out. % with increasing mask number. 
	Black curves give me the measurements for each mask number, shaded bands reflect 
	1-sigma errors from noise and multipole leakage. %, which does not include sample variance from the quadrupole (left) or octopole (right).
	For example, as explained in detail in \Sec{BasicResultsSec}, the quadrupole error bars 
	(left) do {\it not} include quadrupole sample
        variance the way an error bar on a measurement of $C_2$ customarily would, since we are interested in the actual $\alm$-values rather than the underlying power spectrum,
	but they {\it do} include sample variance aliased 
	from higher multipoles.
	Based on our simulations, the coefficient most susceptible to foreground contamination from the
	Galactic plane is $a_{20}$ (top left band), so its mask 6 value is probably closer to the truth than its mask 0 value.
	}
\end{figure}

\subsection{Basic results and their interpretation}

\label{BasicResultsSec}

\begin{table*}[t]
\caption{Measured multipole coefficients in $\mu$K. More coefficients are available online. The relation between these
real-valued spherical harmonic coefficients and the usual complex ones is given in footnote 2. 
} % CMBpolTable
\medskip
\centerline{\label{tabdetectors}
\begin{tabular}{|l|r|r|r|r|r|r|r|r|}
\hline
\hline
\multicolumn{1}{|l|}{} & 
\multicolumn{1}{|c|}{}   & 
\multicolumn{1}{|c|}{Mask 0}&
\multicolumn{3}{|c|}{Mask 5}& 
\multicolumn{3}{|c|}{Mask 6}\\
\cline{3-9}
\multicolumn{1}{|l|}{$\ell$} & 
\multicolumn{1}{|c|}{$m$}   & 
\multicolumn{1}{|c|}{Fit all}&
\multicolumn{1}{|c|}{Fit $0,1,\ell$}& 
\multicolumn{1}{|c|}{Fit $0-3$}& 
\multicolumn{1}{|c|}{Fit $0-5$}& 
\multicolumn{1}{|c|}{Fit $0,1,\ell$}& 
\multicolumn{1}{|c|}{Fit $0-3$}& 
\multicolumn{1}{|c|}{Fit $0-5$}\\
\hline
 2 &-2 & -21.29$\pm$   0.20 & -20.48$\pm$   1.21 & -20.56$\pm$   1.22 & -20.63$\pm$   1.22 & -24.07$\pm$   2.49 & -24.28$\pm$   2.49 & -24.36$\pm$   2.53 \\
 2 &-1 &   5.94$\pm$   0.20 &   5.76$\pm$   0.23 &   5.76$\pm$   0.23 &   5.76$\pm$   0.23 &   5.89$\pm$   0.40 &   5.90$\pm$   0.40 &   5.93$\pm$   0.40 \\
 2 & 0 &  10.61$\pm$   0.22 &  13.64$\pm$   1.11 &  13.69$\pm$   1.11 &  13.86$\pm$   1.12 &  14.04$\pm$   2.29 &  14.10$\pm$   2.30 &  14.32$\pm$   2.34 \\
 2 & 1 &  -8.30$\pm$   0.20 &  -8.21$\pm$   0.26 &  -8.21$\pm$   0.26 &  -8.20$\pm$   0.26 &  -9.14$\pm$   0.59 &  -9.15$\pm$   0.59 &  -9.08$\pm$   0.60 \\
 2 & 2 & -19.39$\pm$   0.20 & -20.99$\pm$   1.48 & -20.97$\pm$   1.48 & -21.16$\pm$   1.49 & -17.81$\pm$   3.10 & -17.70$\pm$   3.11 & -18.53$\pm$   3.18 \\
\hline
 3 &-3 &  40.56$\pm$   0.20 &  43.17$\pm$   1.49 &  43.19$\pm$   1.50 &  43.32$\pm$   1.51 &  41.84$\pm$   3.00 &  41.94$\pm$   3.00 &  42.15$\pm$   3.06 \\
 3 &-2 &   2.54$\pm$   0.20 &   2.29$\pm$   0.27 &   2.29$\pm$   0.27 &   2.28$\pm$   0.27 &   2.87$\pm$   0.50 &   2.86$\pm$   0.50 &   2.84$\pm$   0.50 \\
 3 &-1 &   1.00$\pm$   0.20 &   2.91$\pm$   1.03 &   2.89$\pm$   1.03 &   2.78$\pm$   1.04 &   2.55$\pm$   2.10 &   2.50$\pm$   2.10 &   1.86$\pm$   2.14 \\
 3 & 0 &  -6.47$\pm$   0.22 &  -6.34$\pm$   0.29 &  -6.34$\pm$   0.29 &  -6.36$\pm$   0.29 &  -7.24$\pm$   0.65 &  -7.23$\pm$   0.65 &  -7.28$\pm$   0.66 \\
 3 & 1 & -12.90$\pm$   0.20 & -16.45$\pm$   1.21 & -16.49$\pm$   1.21 & -16.72$\pm$   1.22 & -12.05$\pm$   2.47 & -12.19$\pm$   2.48 & -12.95$\pm$   2.55 \\
 3 & 2 &  30.37$\pm$   0.20 &  30.76$\pm$   0.26 &  30.76$\pm$   0.26 &  30.76$\pm$   0.26 &  31.53$\pm$   0.72 &  31.51$\pm$   0.72 &  31.45$\pm$   0.73 \\
 3 & 3 & -19.32$\pm$   0.20 & -19.02$\pm$   1.41 & -18.98$\pm$   1.41 & -18.97$\pm$   1.42 & -20.76$\pm$   2.95 & -20.63$\pm$   2.96 & -20.94$\pm$   3.04 \\
\hline
 4 &-4 &  -9.69$\pm$   0.20 & -12.12$\pm$   1.51 &                    & -12.49$\pm$   1.52 &  -9.27$\pm$   3.07 &                    & -10.47$\pm$   3.10 \\
 4 &-3 & -29.85$\pm$   0.20 & -29.92$\pm$   0.33 &                    & -29.93$\pm$   0.33 & -29.98$\pm$   0.69 &                    & -30.00$\pm$   0.69 \\
 4 &-2 &   5.24$\pm$   0.20 &   4.68$\pm$   1.03 &                    &   4.82$\pm$   1.04 &   7.65$\pm$   2.07 &                    &   7.97$\pm$   2.09 \\
 4 &-1 &  11.38$\pm$   0.20 &  11.66$\pm$   0.28 &                    &  11.66$\pm$   0.28 &  11.32$\pm$   0.58 &                    &  11.26$\pm$   0.58 \\
 4 & 0 &  18.81$\pm$   0.23 &  15.95$\pm$   1.10 &                    &  15.67$\pm$   1.11 &  15.47$\pm$   2.20 &                    &  14.87$\pm$   2.23 \\
 4 & 1 &  -8.95$\pm$   0.20 &  -9.04$\pm$   0.32 &                    &  -9.05$\pm$   0.32 &  -7.57$\pm$   0.88 &                    &  -7.66$\pm$   0.89 \\
 4 & 2 &  11.75$\pm$   0.20 &  13.09$\pm$   1.26 &                    &  13.20$\pm$   1.27 &  10.98$\pm$   2.52 &                    &  11.45$\pm$   2.57 \\
 4 & 3 &  10.23$\pm$   0.20 &   9.89$\pm$   0.25 &                    &   9.89$\pm$   0.25 &   8.42$\pm$   0.74 &                    &   8.45$\pm$   0.75 \\
 4 & 4 &  -2.46$\pm$   0.20 &  -0.44$\pm$   1.53 &                    &  -0.44$\pm$   1.54 &   3.64$\pm$   3.10 &                    &   3.67$\pm$   3.14 \\
\hline
 5 &-5 &  25.01$\pm$   0.20 &  26.42$\pm$   1.61 &                    &  26.53$\pm$   1.61 &  26.81$\pm$   3.20 &                    &  27.09$\pm$   3.22 \\
 5 &-4 &  11.98$\pm$   0.20 &  11.49$\pm$   0.34 &                    &  11.48$\pm$   0.34 &  11.95$\pm$   0.77 &                    &  11.84$\pm$   0.77 \\
 5 &-3 &   7.70$\pm$   0.20 &   5.98$\pm$   1.21 &                    &   5.72$\pm$   1.22 &   7.05$\pm$   2.34 &                    &   6.19$\pm$   2.38 \\
 5 &-2 &   2.46$\pm$   0.20 &   2.78$\pm$   0.31 &                    &   2.78$\pm$   0.31 &   2.00$\pm$   0.66 &                    &   2.04$\pm$   0.66 \\
 5 &-1 &   5.31$\pm$   0.21 &   3.46$\pm$   1.01 &                    &   3.49$\pm$   1.01 &   4.31$\pm$   2.01 &                    &   4.46$\pm$   2.02 \\
 5 & 0 &  15.61$\pm$   0.23 &  15.34$\pm$   0.34 &                    &  15.35$\pm$   0.34 &  16.46$\pm$   0.83 &                    &  16.44$\pm$   0.83 \\
 5 & 1 &  30.49$\pm$   0.21 &  34.32$\pm$   1.18 &                    &  34.40$\pm$   1.18 &  31.00$\pm$   2.30 &                    &  31.23$\pm$   2.35 \\
 5 & 2 & -11.21$\pm$   0.20 & -11.76$\pm$   0.32 &                    & -11.78$\pm$   0.32 & -12.68$\pm$   0.95 &                    & -12.73$\pm$   0.96 \\
 5 & 3 &  26.11$\pm$   0.20 &  25.84$\pm$   1.15 &                    &  25.82$\pm$   1.16 &  26.67$\pm$   2.28 &                    &  26.81$\pm$   2.33 \\
 5 & 4 &  -6.84$\pm$   0.20 &  -7.14$\pm$   0.28 &                    &  -7.15$\pm$   0.28 &  -5.49$\pm$   0.83 &                    &  -5.55$\pm$   0.83 \\
 5 & 5 &  14.00$\pm$   0.20 &  12.74$\pm$   1.53 &                    &  12.76$\pm$   1.54 &  14.22$\pm$   3.10 &                    &  14.29$\pm$   3.12 \\
  \hline
  \hline					 	 
\end{tabular} 
}
\end{table*}

\begin{table*}[t]
\caption{Same as previous table, but for the 3-year WMAP data.
} % CMBpolTable
\medskip
\centerline{\label{tabdetectors}
\begin{tabular}{|l|r|r|r|r|r|r|r|r|}
\hline
\hline
\multicolumn{1}{|l|}{} & 
\multicolumn{1}{|c|}{}   & 
\multicolumn{1}{|c|}{Mask 0}&
\multicolumn{3}{|c|}{Mask 5}& 
\multicolumn{3}{|c|}{Mask 6}\\
\cline{3-9}
\multicolumn{1}{|l|}{$\ell$} & 
\multicolumn{1}{|c|}{$m$}   & 
\multicolumn{1}{|c|}{Fit all}&
\multicolumn{1}{|c|}{Fit $0,1,\ell$}& 
\multicolumn{1}{|c|}{Fit $0-3$}& 
\multicolumn{1}{|c|}{Fit $0-5$}& 
\multicolumn{1}{|c|}{Fit $0,1,\ell$}& 
\multicolumn{1}{|c|}{Fit $0-3$}& 
\multicolumn{1}{|c|}{Fit $0-5$}\\
\hline
 2 &-2 & -24.29$\pm$   0.20 & -23.06$\pm$   1.21 & -23.14$\pm$   1.22 & -23.20$\pm$   1.22 & -25.88$\pm$   2.49 & -26.09$\pm$   2.49 & -26.20$\pm$   2.53 \\
 2 &-1 &   6.83$\pm$   0.20 &   6.70$\pm$   0.23 &   6.69$\pm$   0.23 &   6.69$\pm$   0.23 &   6.65$\pm$   0.40 &   6.66$\pm$   0.40 &   6.69$\pm$   0.40 \\
 2 & 0 &   3.22$\pm$   0.22 &   3.88$\pm$   1.11 &   3.94$\pm$   1.11 &   4.11$\pm$   1.12 &   3.79$\pm$   2.29 &   3.86$\pm$   2.30 &   4.14$\pm$   2.34 \\
 2 & 1 &   0.37$\pm$   0.20 &   0.51$\pm$   0.26 &   0.51$\pm$   0.26 &   0.52$\pm$   0.26 &  -0.62$\pm$   0.59 &  -0.62$\pm$   0.59 &  -0.54$\pm$   0.60 \\
 2 & 2 & -21.06$\pm$   0.20 & -19.54$\pm$   1.48 & -19.52$\pm$   1.48 & -19.72$\pm$   1.49 & -15.31$\pm$   3.10 & -15.18$\pm$   3.11 & -16.11$\pm$   3.18 \\
\hline
 3 &-3 &  45.13$\pm$   0.20 &  47.03$\pm$   1.49 &  47.06$\pm$   1.50 &  47.24$\pm$   1.51 &  44.50$\pm$   3.00 &  44.59$\pm$   3.00 &  44.97$\pm$   3.06 \\
 3 &-2 &   2.36$\pm$   0.20 &   2.11$\pm$   0.27 &   2.11$\pm$   0.27 &   2.10$\pm$   0.27 &   2.62$\pm$   0.50 &   2.62$\pm$   0.50 &   2.59$\pm$   0.50 \\
 3 &-1 &   8.44$\pm$   0.20 &  10.58$\pm$   1.03 &  10.55$\pm$   1.03 &  10.46$\pm$   1.04 &   8.72$\pm$   2.10 &   8.66$\pm$   2.10 &   8.07$\pm$   2.14 \\
 3 & 0 &  -8.45$\pm$   0.22 &  -8.27$\pm$   0.29 &  -8.27$\pm$   0.29 &  -8.28$\pm$   0.29 &  -9.52$\pm$   0.65 &  -9.50$\pm$   0.65 &  -9.54$\pm$   0.66 \\
 3 & 1 & -15.88$\pm$   0.20 & -15.97$\pm$   1.21 & -16.00$\pm$   1.21 & -16.24$\pm$   1.22 & -11.94$\pm$   2.47 & -12.03$\pm$   2.48 & -12.86$\pm$   2.55 \\
 3 & 2 &  32.65$\pm$   0.20 &  32.96$\pm$   0.26 &  32.96$\pm$   0.26 &  32.96$\pm$   0.26 &  33.71$\pm$   0.72 &  33.70$\pm$   0.72 &  33.61$\pm$   0.73 \\
 3 & 3 & -16.76$\pm$   0.20 & -18.59$\pm$   1.41 & -18.55$\pm$   1.41 & -18.52$\pm$   1.42 & -20.71$\pm$   2.95 & -20.60$\pm$   2.96 & -20.79$\pm$   3.04 \\
\hline
 4 &-4 & -12.24$\pm$   0.20 & -14.39$\pm$   1.51 &                    & -14.81$\pm$   1.52 & -11.83$\pm$   3.07 &                    & -13.10$\pm$   3.10 \\
 4 &-3 & -27.82$\pm$   0.20 & -27.77$\pm$   0.33 &                    & -27.79$\pm$   0.33 & -27.60$\pm$   0.69 &                    & -27.62$\pm$   0.69 \\
 4 &-2 &   5.74$\pm$   0.20 &   4.79$\pm$   1.03 &                    &   4.92$\pm$   1.04 &   7.10$\pm$   2.07 &                    &   7.43$\pm$   2.09 \\
 4 &-1 &   9.79$\pm$   0.20 &   9.99$\pm$   0.28 &                    &  10.00$\pm$   0.28 &   9.89$\pm$   0.58 &                    &   9.84$\pm$   0.58 \\
 4 & 0 &  17.25$\pm$   0.23 &  16.71$\pm$   1.10 &                    &  16.43$\pm$   1.11 &  16.62$\pm$   2.20 &                    &  16.02$\pm$   2.23 \\
 4 & 1 & -14.43$\pm$   0.20 & -14.63$\pm$   0.32 &                    & -14.64$\pm$   0.32 & -12.94$\pm$   0.88 &                    & -13.02$\pm$   0.89 \\
 4 & 2 &  16.33$\pm$   0.20 &  15.01$\pm$   1.26 &                    &  15.12$\pm$   1.27 &  12.45$\pm$   2.52 &                    &  12.89$\pm$   2.57 \\
 4 & 3 &  12.17$\pm$   0.20 &  11.91$\pm$   0.25 &                    &  11.90$\pm$   0.25 &   9.89$\pm$   0.74 &                    &   9.92$\pm$   0.75 \\
 4 & 4 &  -5.72$\pm$   0.20 &  -3.22$\pm$   1.53 &                    &  -3.23$\pm$   1.54 &   1.14$\pm$   3.10 &                    &   1.19$\pm$   3.14 \\
\hline
 5 &-5 &  27.10$\pm$   0.20 &  28.39$\pm$   1.61 &                    &  28.52$\pm$   1.61 &  28.69$\pm$   3.20 &                    &  29.01$\pm$   3.22 \\
 5 &-4 &  11.39$\pm$   0.20 &  10.89$\pm$   0.34 &                    &  10.88$\pm$   0.34 &  11.41$\pm$   0.77 &                    &  11.28$\pm$   0.77 \\
 5 &-3 &   3.89$\pm$   0.20 &   2.69$\pm$   1.21 &                    &   2.40$\pm$   1.22 &   4.57$\pm$   2.34 &                    &   3.63$\pm$   2.38 \\
 5 &-2 &   2.56$\pm$   0.20 &   2.89$\pm$   0.31 &                    &   2.90$\pm$   0.31 &   2.21$\pm$   0.66 &                    &   2.26$\pm$   0.66 \\
 5 &-1 &   0.90$\pm$   0.21 &  -1.15$\pm$   1.01 &                    &  -1.14$\pm$   1.01 &   0.98$\pm$   2.01 &                    &   1.09$\pm$   2.02 \\
 5 & 0 &  15.78$\pm$   0.23 &  15.45$\pm$   0.34 &                    &  15.45$\pm$   0.34 &  16.98$\pm$   0.83 &                    &  16.94$\pm$   0.83 \\
 5 & 1 &  34.10$\pm$   0.21 &  34.50$\pm$   1.18 &                    &  34.58$\pm$   1.18 &  31.71$\pm$   2.30 &                    &  31.94$\pm$   2.35 \\
 5 & 2 &  -9.74$\pm$   0.20 & -10.19$\pm$   0.32 &                    & -10.21$\pm$   0.32 & -11.06$\pm$   0.95 &                    & -11.10$\pm$   0.96 \\
 5 & 3 &  24.12$\pm$   0.20 &  25.59$\pm$   1.15 &                    &  25.55$\pm$   1.16 &  26.41$\pm$   2.28 &                    &  26.53$\pm$   2.33 \\
 5 & 4 &  -5.72$\pm$   0.20 &  -5.95$\pm$   0.28 &                    &  -5.97$\pm$   0.28 &  -3.74$\pm$   0.83 &                    &  -3.83$\pm$   0.83 \\
 5 & 5 &  13.47$\pm$   0.20 &  12.37$\pm$   1.53 &                    &  12.41$\pm$   1.54 &  13.25$\pm$   3.10 &                    &  13.37$\pm$   3.12 \\
  \hline
  \hline					 	 
\end{tabular} 
}
\end{table*}

For our basic results, we work at HealPix pixel level $nside=16$ and use 
the $\C$-matrix corresponding to noise and CMB measurements by the WMAP team \cite{bennett03b}, with 
$C_\l$ in \eq{Ceq} given by be the best fit $\Lambda$CDM model from \cite{spergel03}.
As detailed below, our low-$\ell$ results are quite insensitive to these choices.

\Fig{tornadoFig} shows the result of applying our method to this data to measure the 
components of the quadrupole and octopole. The results are plotted as a function of 
Galatic cut, with the thin black lines showing the measurements $\almhat$ and the shaded 
bands illustrating the $1\sigma$ error range $\almhat\pm\Delta\almhat$. These plotted 
error bars are $\Delta\widehat{a}_{\l_j m_j}=\Delta\widehat{a}_j=\SS_{jj}^{1/2}$. 
The results for more multipoles are listed in Table 1. The measurements using the full 
sky (Mask 0) match those reported in \cite{tegmark03,smalluniverse}. Table 2 shows the corresponding results for the 3-year WMAP data, which was released after the original submission of this paper.
The results are quite similar --- we discuss the differences below in \Sec{AlignmentSec} and \Sec{ConcSec}.\footnote{We foreground-cleaned the 3 year WMAP data with the
exact same TOH algorithm \cite{tegmark03} 
as was used for the 1-year data, and the interested reader can download the cleaned map from\\
\url{http://space.mit.edu/home/tegmark/wmap.html}.
}
The quadrupole moments were computed by taking $\a$ to include all multipole 
coefficients with $\l\le 2$. It is important to always include the monopole $(\l=0)$
and dipole $(\l=1)$ so that our method makes the measurements that we are interested in
(in this case $\l=2$) completely independent of these totally unknown 
quantities (the maps released by the WMAP team have already been approximately purged of a 
3K 
monopole corresponding to the mean CMB temperature and a 
6$\,$mK 
dipole corresponding to Earth's motion).
Similarly, the octopole measurements were computed by taking $\a$ to include all multipole 
coefficients with $\l=3$ and $\l\le 1$ (We will use the alternative $\l\le 3$ option
in \Sec{AlignmentSec} where we study quadrupole-octopole alignment and want uncorrelated measurements
of the two.)

Note that the error bars $\Delta\almhat$ on, say, the quadrupole coefficients 
shown in \fig{tornadoFig} do {\it not} include quadrupole sample
variance the way an error bar on a measurement of $C_2$ customarily would.
This is why they are seen to be so tiny for Mask 0, reflecting only the detector noise contribution. 
The error bar $\Delta\almhat$ simply reflects the uncertainty in our measurement 
of the coefficient $\almhat$ on the sky seen from our particular vantage point in space.
These coefficients vary randomly with a variance $C_2$ between widely separated Hubble volumes,
and this quadrupole sample variance therefore enters only in the next data analysis step
where our five locally measured quadrupole coefficients are used to make inferences 
about the value of $C_2$.

As mentioned in \Sec{MethodSec}, however, the error bars $\Delta\almhat$ {\it do} include
sample variance from the other multipoles that were excluded from the $\a$-vector,
\ie, $\l\ge 3$ for our quadrupole example.
This is why the error bars in \fig{tornadoFig} are seen to flare up dramatically 
towards the right: as large portions of the sky get masked out, the severely 
broken orthogonality between the spherical harmonics makes higher multipoles 
contribute to quadrupole estimators.
Since the values of these higher multipole coefficients
are unknown to us and are not measured by the method in this case, 
their contribution is counted as noise --- which can be quantified since their variance 
$\Cl$ is known.
Since our method tries to minimize the error bars, it makes the best tradeoffs it can to
minimize the net effect of such leakage from non-included multipoles. 
For multipoles included in $\a$, in contrast, our method forces the leakage to be exactly zero, 
at the cost of larger error bars and more leakage from non-included  multipoles.

The disk-like geometry of our Galaxy provides some intuition for the behavior of the error bars.
Because the Galaxy cuts are approximately symmetric under reflection (parity-even),
different $\l$-values couple mainly if they are separated by an even number.
To the extent that the Galaxy cut is azimuthally symmetric (a crude approximation at best),
different $m$-values do not couple.
For example, the quadrupole estimator $\widehat{a}_{21}$ therefore picks up ``noise'' mainly
from $a_{41}$, $a_{61}$, {\etc}
Moreover, since spherical harmonics with high $|m|$ live mainly near the Galactic equator
and rapidly approach zero towards the Galactic poles, they are the ones that 
suffer most as the cut is increased and therefore have the largest error bars in 
Table 1.

The above-mentioned sample variance considerations also shed light on how the method is affected
by changes in the $\C$-matrix, \ie, on how robust our method is to assumptions about unwanted noise and CMB signals.
Since the method is required to faithfully (without bias) recover the multipoles $\a$ whatever they are,
one would intuitively expect that $\W$ and therefore $\ahat$
are independent of contributions from signal and noise in these multipoles to the 
$\C$-matrix. 
With some matrix algebra, one can prove that that this is indeed the case, specifically that
adding to $\C$ a contribution of the form 
$\Y\SSS\Y^t$ for some matrix $\SSS$ leaves $\W$ and $\ahat$ unchanged and simply 
increases $\SS$ by $\SSS$. For example, if we use our method to measure the five quadrupole coefficients,
the result $\ahat$ will be independent of our assumptions about the true value of $C_2$.

\subsection{Simulations with foregrounds}

The key remaining issue is how to quantify the contribution of residual foreground contamination and how to optimize the
Galaxy cut to minimize this contribution. We will now address this issue with simulations.

The residual foreground contamination present after foreground cleaning can be quite important.
All five linearly cleaned WMAP-maps used in the literature have their problems.
The WMAP team ILC map \cite{bennett03b} comes with the disclaimer that it should not be used for scientific 
analysis --- needless to say, the science questions are so interesting that this has not stopped 
large numbers of groups from using it anyway, particularly when all-sky coverage was helpful.
The TOH cleaned WMAP map \cite{tegmark03} has the advantage of having lower total noise 
and residual foregrounds, but like the ILC map, it lacks error bars accurately quantifying these residuals, as do all other 
foreground-cleaned maps published to date.
The SMICA map \cite{Patanchon04} is generated with a method very similar that used for the TOH map \cite{wiener,tegmark03}, and the result is 
indeed encouragingly similar \cite{Patanchon04}.
The WI-FIT map of \cite{Hansen06} is also closely related (since this method cleans with linear combinations of
linear combinations (pairwise differences) of maps, it too is a linear and WMAP-internal technique just like the ILC, TOH and SMICA methods --- the 
only fundamental difference between the four methods is that the approximate variance minimizes minimization is performed in 
pixel space for ILC, in (masked) spherical harmonic space for TOH and SMICA and in wavelet space for WI-FIT (and optionally SMICA).
Finally, using external template maps alone \cite{bennett03b} suffers from both 
dubious extrapolation and from the problem that they (\eg, the Haslam-based synchrotron map)
have striping or other systematic errors at a substantially larger level than
the WMAP maps, and that these errors will propagate into the cleaned map,
perhaps making it less clean that it was to start with in the most systematics-affected modes
--- template-cleaned residual WMAP foregrounds have never been accurately quantified in the
literature.

A useful recent approach has been to quantify ILC errors by reapplying the ILC cleaning algorithm
to simulated sky maps where the correct answer is known and the residuals can thus be 
computed \cite{eriksen0508196}. 
This has confirmed that the residual foreground contribution is substantial.
We will adopt a similar procedure for the TOH map, but with an important difference that makes our approach more conservative.

\subsubsection{Mock foregrounds}

It is absolutely crucial to use a foreground
model that does not artificially mimic the cleaning algorithm. For example, as acknowledged
in Eriksen {\etal} (2005) \cite{eriksen0508196}, 
their simulations underestimate the residual foregrounds
by assuming that the frequency dependence of each physical component is the same in all directions in the
sky --- this assumption of perfect frequency coherence \cite{foregpars}
is empirically known to be incorrect for synchrotron radiation (see, \eg, \cite{astro-ph/0511384} for a recent treatment), and allows
the simulated foregrounds to be perfectly removed by the cleaning method.
To avoid reaching overly optimistic conclusions, we therefore
make simulated foreground templates at the five WMAP frequencies in a different way: 
straight from data by subtracting  the TOH CMB maps from the five WMAP frequency channels.
The resulting five foreground maps that we use are shown in  in \fig{mapFig} of \cite{tegmark03}.
This procedure is probably overly conservative because some foregrounds and noise get double-counted: 
for instance, our K-band ``foreground'' map includes detector noise from 
all five WMAP frequencies because it was constructed by subtracting the TOH map (with its noise) 
from the K-band observations.

\subsubsection{Mock CMB} 

We generate a simulated CMB-only sky map (\fig{mapFig}, top right) 
by using the HEALPix software and the best fit WMAP power 
spectrum from \cite{spergel03}, smoothed to have the same angular resolution as the TOH map
(corresponding to the beam of the W-band, the WMAP channel with the highest resolution).
As input to the cleaning algorithm, we also smoothed it to the four lower resolutions corresponding to the 
K-, Ka-, Q- and W-band beams \cite{Page03}.

\subsubsection{Mock detector noise}
 
We construct simulated WMAP noise maps at the five channels by assuming that the noise is
Gaussian and uncorrelated both between different pixels and between different channels, which is a good approximation
according to \cite{bennett03b}.
We set the noise amplitude in each map so that the rms fluctuation levels match those estimated from the 
WMAP team (Table 5 in \cite{Hinshaw03}). We confirmed that this noise normalization matched both the rms in 
various same-frequency difference maps such as $W=(W1-W2+W3-W4)/4$ and the total power spectrum normalization 
at very high $\l$ where CMB and foregrounds are negligible. 
Specifically, we used rms noise values $\N_{ii}^{1/2}$ of 
$150\uK$, 
$132\uK$, 
$122\uK$, 
$149\uK$ and
$179\uK$ for K, Ka, Q, V and W, respectively.
This corresponds to an $\l$-independent noise power spectrum of order
$C_\l\sim 0.1\uK^2$ in each channel, \ie, orders of magnitude below the large-scale CMB signal 
(\cite{tegmark03} estimates $C_2\sim 200\uK^2$ and $C_2\sim 500\uK^2$),
so the details of our noise modeling have no impact whatsoever on our results --- rather, our uncertainties
are dominated by sample variance and residual foregrounds.
We tried both crude noise maps with equal variance in all pixels and more accurate ones 
with the variance modulated in inverse proportion to the number of WMAP observations of each pixel.
As expected from the fact that noise is negligible at the low $\l$-values of interest here, 
these two approaches give for all practical purposes indistinguishable results --- we adopt the former
for simplicity.

\subsubsection{Residuals} 

For each of the five WMAP channels, we sum the mock CMB, noise and foreground maps described above.
We then apply the TOH foreground cleaning method to these five input maps exactly as described
in \cite{tegmark03} to obtain a single  foreground-cleaned map.
Finally, we subtract the ``true'' simulated CMB sky map from the cleaned map, obtaining the
residual map shown in \fig{mapFig} (bottom left). This figure visually illustrates that
the TOH method provides an unbiased recovery of the CMB, since the residual map 
shows no patterns whatsoever in common with the input CMB map (upper right).
Comparing the residual map with the V-band map (top left) also illustrates that the
TOH method reduces the levels of both foregrounds and detector noise.
Nonetheless, residual foregrounds are clearly visible, particularly near the Galactic plane
in the parts that the ``junk  map'' (bottom right) suggests are highly contaminated.

If we knew that this was the {\it actual} residual in our TOH map, we could of course
eliminate the foreground problem completely by subtract it out.
This is unfortunately not the case. To be conservative, will make no attempt to further clean our input map, 
and will merely use the residual map to quantify the errors in our multipole measurements.

What does the residual map depend on?
By repeating the entire procedure with independent CMB simulations, we found that the 
residual map was rather independent of what CMB map was used. This is expected, since 
the TOH method is completely CMB-independent except for the small effect that CMB 
sample variance can have on the computation of optimal weights for the cleaning process.
Repeating the entire procedure with independent noise simulations alters the
residual map mainly on the very smallest angular scales.
The large-scale patterns visible in \fig{mapFig} (bottom left) are therefore
determined mainly by properties of the foregrounds themselves. 
For example, foregrounds whose spectral index varies notably across the sky cannot be
perfectly subtracted out by the TOH method.
Since these large-scale residual patterns are all that matter for our present focus on low 
$\l$, there is no point in making more than one simulation of our type, \ie, it is 
pointless to make an ensemble of simulations with different CMB and noise realizations.

\begin{figure} 
\centerline{\epsfxsize=\figsize\epsffile{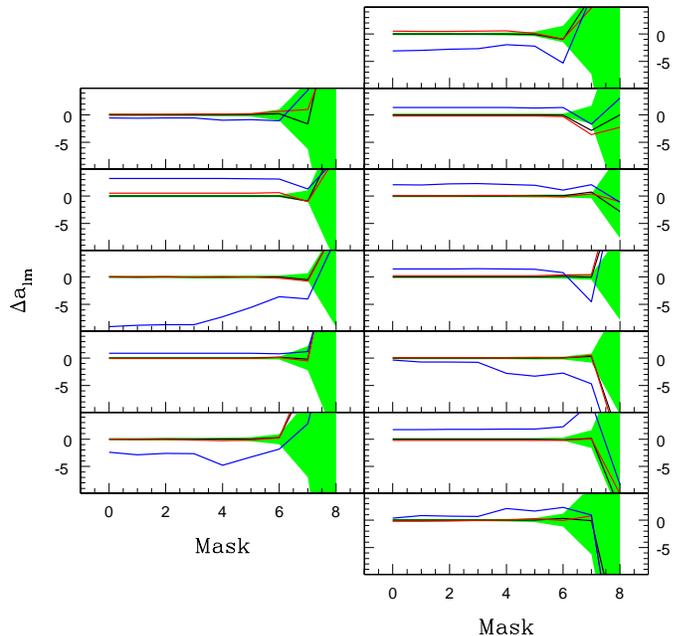}}
\caption[1]{\label{foregFig}\footnotesize%
The left and right panels show how the
measurement errors $\Delta\alm\equiv\widehat{a}_{\l m}-\alm^{\rm true}$ in $\mu$K
for the quadrupole (left) and 
octopole (right) extracted from a simulated map change with 
increasing Galactic cut. The three curves in each panel correspond to 
errors from CMB multipole leakage alone 
(black)
CMB plus detector noise 
(red)
and CMB plus noise plus residual foregrounds 
(blue)
The shaded/green areas show the 
$1\sigma$ error bars corresponding to leakage and noise.
Note that these foreground-related errors are typically an order of magnitude smaller than the expected CMB signal
(the $\Lambda$CDM concordance model gives $C_2^{1/2} = 30\mu$K and $C_3^{1/2} = 20\mu$K), and that they are even more subdominant for power spectrum estimation where they 
add only in quadrature.
}
\end{figure}

\subsection{Optimizing the Galaxy cut}

To quantify the effect of residual foregrounds on our multipole measurements, we
repeat the analysis from \Sec{ApplicationSec} using 
our simulated maps. Specifically, we measure the $\alm$-coefficients from 
the simulated CMB-only map, from a map with CMB and noise 
(made without introducing any foregrounds before the cleaning process) and from the cleaned map including 
CMB, noise and foregrounds. The results are shown in 
\Fig{foregFig} with the correct (simulated) $\alm$-values subtracted off.

The fact that the black and red curves are almost identical 
confirms the above-mentioned claim that WMAP detector noise makes essentially
no difference on these large angular scales.
The fact that the red curves agree well with the shaded 
regions confirms that our method and our software implementing it are working as
they should, \ie, that the multipoles are faithfully recovered
with accuracy consistent with the predicted
noise$+$leakage error bars $\Delta\almhat$.

The key new information in \fig{foregFig} is contained in the 
blue curves, which include errors caused by residual foregrounds.
These curves reveal two interesting facts: 
\begin{enumerate}
\item Residual foreground contamination near the Galactic plane afflicts the $a_{20}$-component at a level around ten $\uK$.
\item Residual foreground contamination at higher latitudes appears to afflict most coefficients at the level of a few $\uK$.
\end{enumerate}

The fact that $a_{20}$ is special follows from the well-known fact that this is the component most similar in shape to the 
Galactic emission, as pointed out already in the COBE/DMR analysis \cite{smoot92}.
Specifically, in the very crude approximation that the Galactic plane has both parity and azimuthal symmetry, 
it would contaminate only the multiple coefficients $a_{20}$, $a_{40}$, $a_{60}$, {\etc}
\Fig{foregFig} shows that most of this residual contamination comes from the innermost parts of the plane, so that
cutting out merely the dirtiest 7\% of the sky (Mask 6) cuts the foreground contribution to $a_{20}$ from 
about $9\uK$ to $3\uK$.

In contrast, and in good agreement with these symmetry considerations, none of the other quadrupole or octopole coefficients 
show evidence of contamination from the inner Galactic plane: the foreground contributions shown by the blue curves 
do not tend to shift closer to the zero as the cut is increased up to Mask 6. Beyond this, the leakage errors are seen
to exceed the residual foreground errors. 
This small residual foreground contamination is seen to typically be $1-3\uK$, of similar magnitude to the
$a_{20}$-contamination that remains for Mask 6.

So based on these results, what is the optimal galaxy cut when measuring multipole coefficients?
We have seen that too small a Galaxy cut leads to unneccessarily high foreground contamination,
whereas too large a cut leads to unneccessarily high leakage from other multipoles.
The natural optimum is seen to be Mask 5 or 6, where the typical errors $\Delta\alm$ are of order 
a $\uK$ from leakage 
and 
a few $\uK$ from foregrounds, and we have therefore listed results for both in Table 1. 
For the more aggressive Mask 7, sample variance strongly dominates, and for the less aggressive Mask 4,
the $a_{20}$ foreground contamination is noticeably larger.
If the reader wishes to perform an analysis that it no way involves leakage from the black sheep $a_{20}$, another natural choice is Mask 0.

\subsection{Optimizing the pixel size}

\begin{figure} 
\centerline{\epsfxsize=\figsize\epsffile{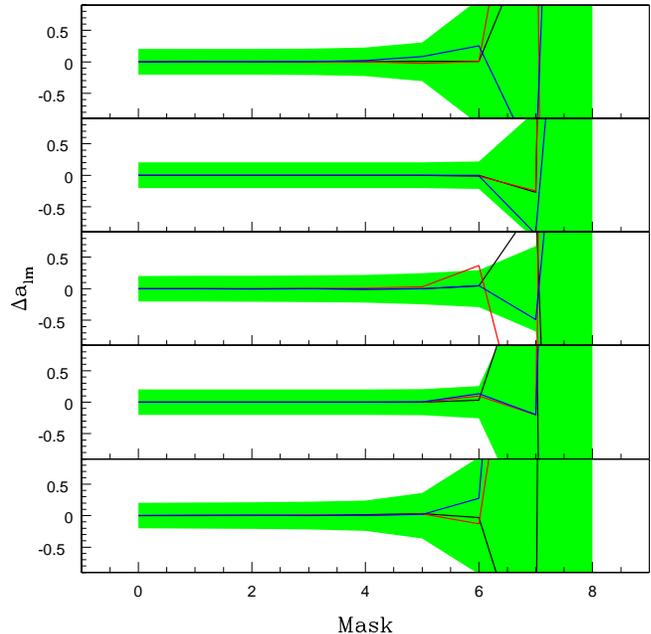}}
\caption[1]{\label{pixelizationFig}\footnotesize%
Effects of pixelization.
The panels show how the measurement errors $\Delta\alm\equiv\widehat{a}_{\l m}-\alm^{\rm true}$ 
in $\mu$K extracted from a simulated map change with 
increasing Galactic cut.
the black, blue and red curves correspond to 
Healpix pixelization at angular resolution level 
$nside=$4, 8 and 16, respectively. 
The shaded/green bands show the 
$1\sigma$ errors from detector noise and leakage.

}
\end{figure}

\begin{figure*} 
\centerline{\epsfxsize=\figsize\epsffile{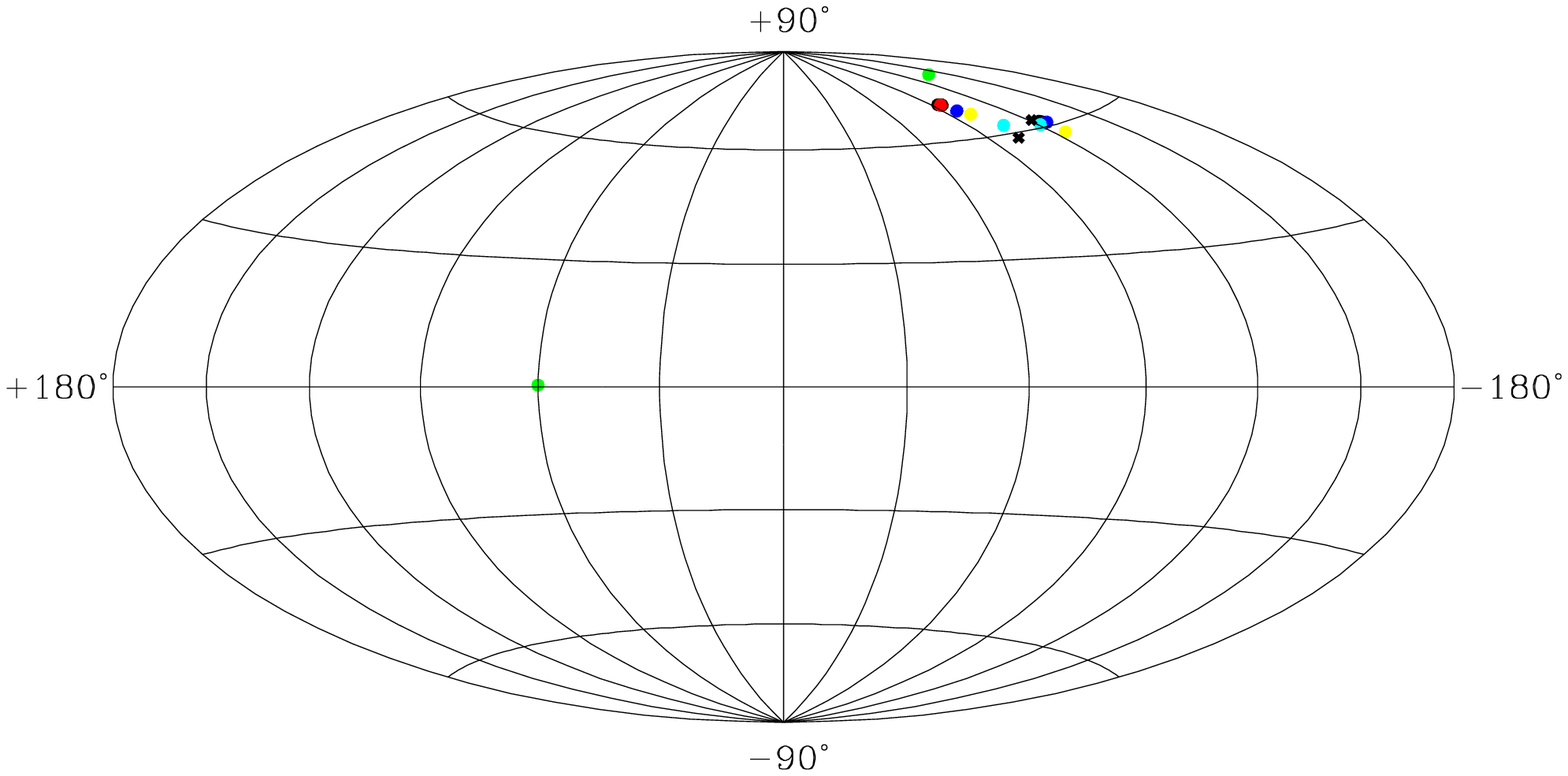}\epsfxsize=\figsize\epsffile{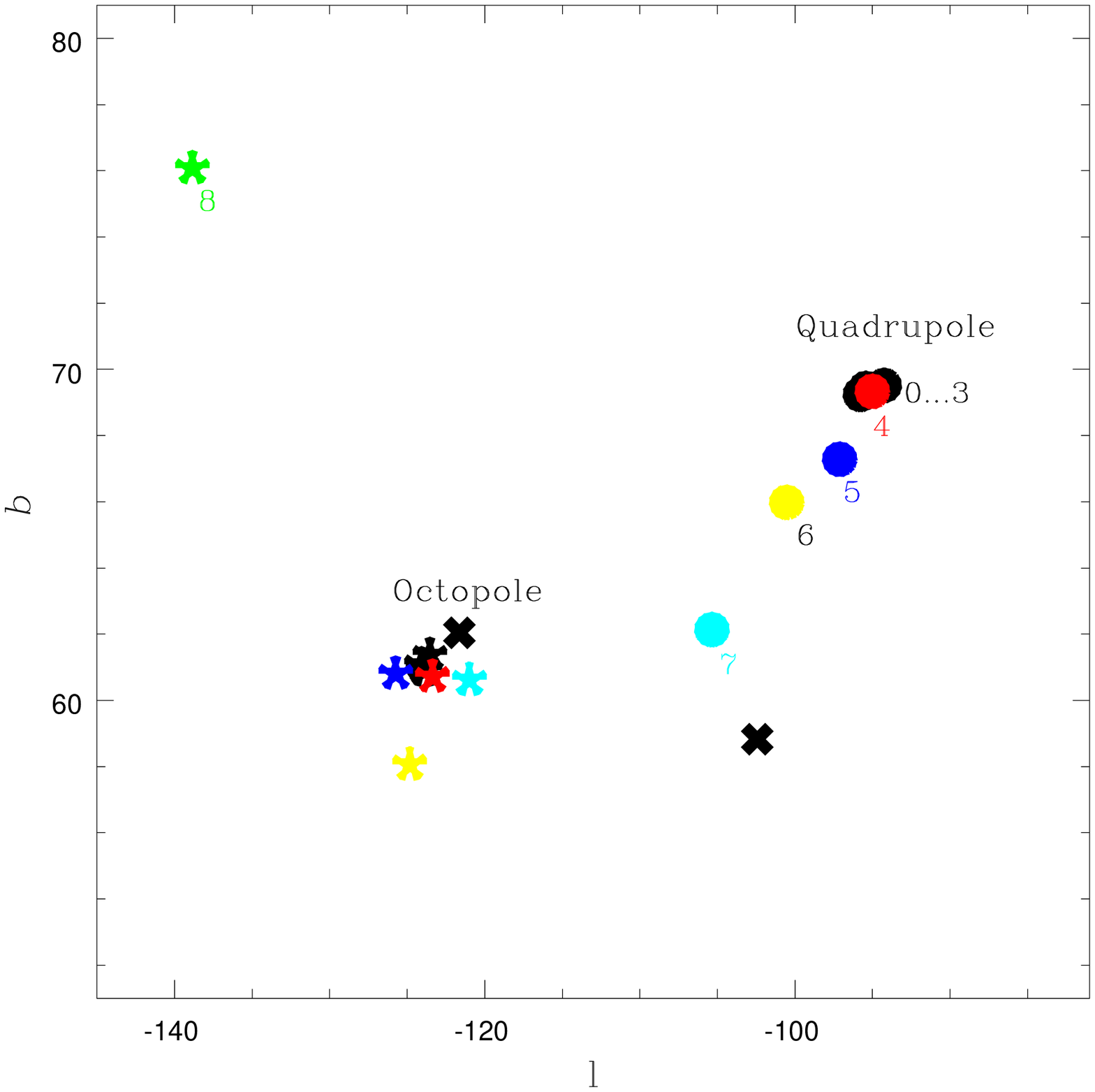}}
\caption[1]{\label{quadrupoleFig}\footnotesize%
Change in the direction of the quadrupole and octopole ``axis of  evil'' directions as a function of the mask number. 
The two crosses correspond to the true quadrupole and octopole axes in our CMB simulation,
chosen to match the WMAP measured values. The departures of the shaded dots from the corresponding crosses reflect the effect of noise, multipole leakage and 
foreground contamination.
}
\end{figure*}

This subsection discusses practical details useful for readers interested in applying this method. Other readers 
may wish to skip straight to \Sec{AlignmentSec}.

Applying equations\eqn{ahatEq} and\eqn{CovEq} using the native WMAP pixel resolution 
($nside=512$) is unfeasible, since it would involve the inversion of an
$N\times N$-dimensional $\C$-matrix with $N$ of order 3 million.
Since we are only interested in the lowest multipoles, there are 
fortunately two simple ways to overcome this problem.
The first approach involves using matrix identities to transform the problem into one involving smaller matrices
(as for the so-called Woodbudy formula).
The second approach, which we used above, is to simply smooth the map onto a
coarser pixelization and correct measurements for this.

\Fig{pixelizationFig} shows the errors on the recovered quadrupole coefficients from our simulated map 
(with neither detector noise nor foregrounds) for pixelizations with $nside$=4, 8 and 16, showing that as long as we work at
resolution 16, pixelization errors are negligibly small compared to noise and sample variance for the masks of interest. 

This pixelization corresponds to averaging the underlying sky map across the area of each pixel.
If the map were subjected to Gaussian smoothing, then the multipoles $\alm$ would be suppressed by the well-known factor 
$e^{\theta^2\l(\l+1)/2}$, where $\theta=$FWHM$/\sqrt{8\ln{2}}$ and FWHM is the full-width half-maximum of the 
Gaussian smoothing kernel. For our adopted resolution level 16, the pixels are approximately squares of side $3.7^\circ$.
A very crude estimate therefore suggests that the corresponding multipole suppression is of order
$e^-(\l/40)^2/2$, \ie, a $0.3\%$ quadrupole suppression and a $2\%$ suppression for $\l=5$.
Since this smoothing effect is negligible compared to the measurement errors caused my leakage and foregrounds, we have not 
corrected for it in Table 1.

In order to quantify the errors due to pixelization, we calculate the five 
components of the quadrupole for different nside$s$ of 4,8 and 16, and also 
studied how these values change as we increase the mask number (or increase 
the Galactic cut). 

Although resolution 8 already
gives us a good estimate of the lower multipoles, we decided (to be conservative) 
to use nside=16 in all calculations done below.

\section{Implications for the quadrupole-octopole alignment}
\label{AlignmentSec}

As a simple example of an application of our results, we use our new multipole measurements to revisit the significance of the
WMAP quadrupole-octopole alignment. 
The two crosses in Figure 5 show the ``axis of evil'' from \cite{tegmark03,smalluniverse}, extracted from a simulated CMB map where we inserted the exact same multipole
and quadrupole coefficients as measured from WMAP.
We then added noise and foregrounds as above and extracted the ``axis of evil'' for our galactic cut. 
Confirming the conclusion of \cite{bielewicz05}, the octopole is seen to be quite robost, whereas the quadrupole moves around somewhat more (it is clearly more fragile 
due to its intrinsically lower amplitude). As seen in Figure 6, this causes the apparent measured alignment be somewhat less significant than the true  
one, but makes no dramatic difference. Similarly, we find that replacing Mask 0 by Mask 6 (the joint quadrupole/octopole fit in Table 1)
degrades the alignment significance only slightly, from a one-in-sixty fluke to a one-in-forty fluke.

\begin{figure} 
\centerline{\epsfxsize=\figsize\epsffile{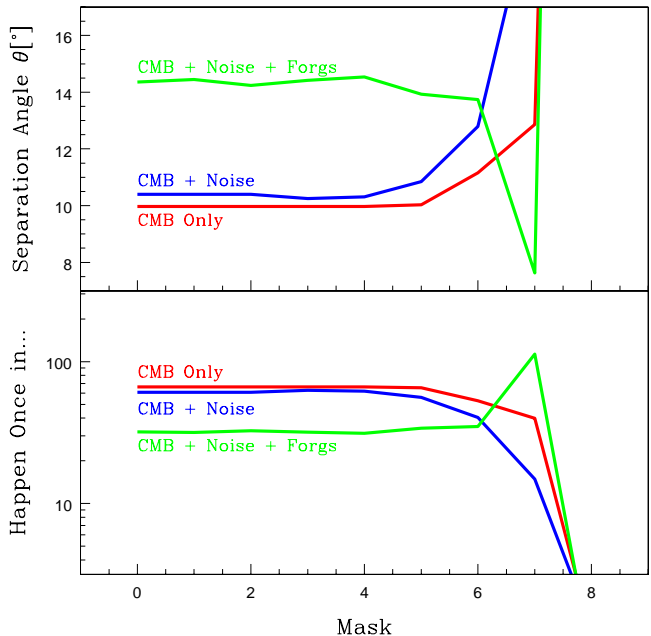}}
\caption[1]{\label{quadrupoleFig}\footnotesize%
The top panel shows the separation angle (in degrees) between the
quadrupole and octopole vectors and how they change as a the mask
number increases. The lower panel shows the probability for this
alignment to happen. In both figures the red, blue and green lines
represent the coefficient values for a map with CMB only, CMB plus noise, 
and CMB plus noise plus residual foreground, respectively.  
}
\end{figure}

\begin{figure*} 
\centerline{\epsfxsize=\figsize\epsffile{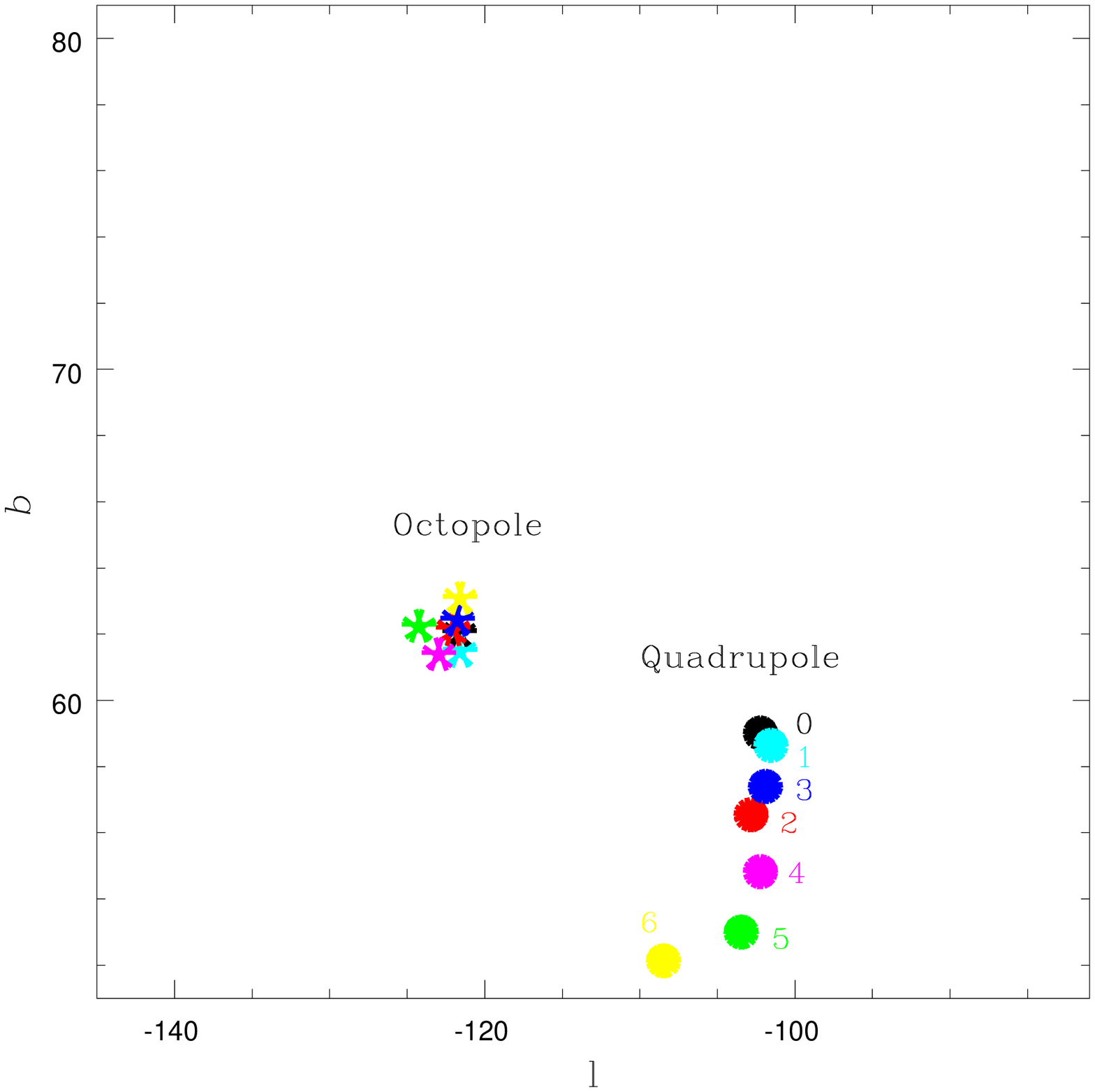}\epsfxsize=\figsize\epsffile{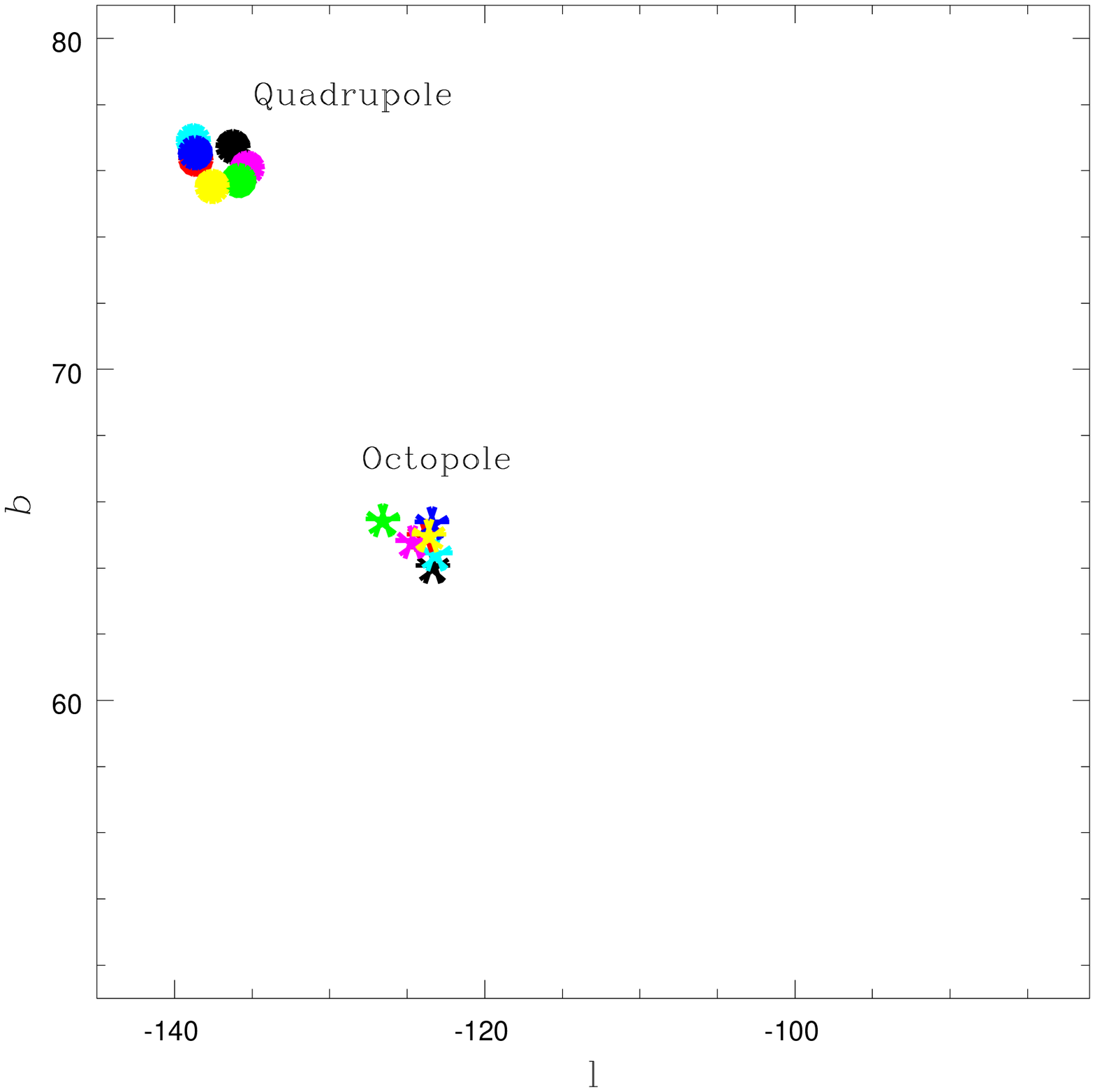}}
\caption[1]{\label{axisofevil}\footnotesize%
Changes in the direction of the preferred quadrupole and octopole axes as a function of mask number 
for the WMAP 1-year (left) and 3-year (right) data. 
In both panels, the quadrupole and octopole results are represented by balls and stars, respectively, 
and the colors black, cyan, red, blue, magenta, green and yellow correspond to the results for
masks from 0 to 6, respectively. The quadrupole axis is seen to be more robust in the 3-year data, suggesting that the lower noise level
may have improved the performance of the foreground cleaning process, reducing the level of residual low-latitude foreground contamination.
}
\end{figure*}

\section{Conclusions}
\label{ConcSec}

We have presented a minimum-variance method for measuring the CMB multipole coefficients
$a_{\l m}$ given anisotropic noise, incomplete sky coverage and foreground contamination.
Our method constitutes lossless data compression 
in the sense that the widely used quadratic estimators of the power spectrum $C_\l$ can 
be computed directly from our $a_{\l m}$-estimators.
We illustrated the method by applying it to the WMAP data.
As the Galactic cut is increased, the error bars $\Delta a_{\l m}$ on low multipoles 
go from being dominated by foregrounds to being dominated by sample variance from other 
multipoles, with the intervening minimum defining the optimal cut.

Because WMAP detector noise is negligible on these large angular scales,
one would not expect the improved sensitivity of second WMAP data release to significantly modify our results {\it per se}. 
Comparing Table 1 with Table 2 shows that the multipoles that change significantly are those most sensitive to foregrounds, notable $a_{20}$, suggesting the interpretation that 
the lower noise levels improved the foreground cleaning. The corresponding mask 6 quadrupole and octopole ``axes of evil'' change from
$(l,b)=(251.5^\circ,52.1^\circ)$ and $(238.4^\circ,63.0^\circ)$ for WMAP1 to 
$(l,b)=(222.4^\circ,75.5^\circ)$ and $(236.4^\circ,64.9^\circ)$ for WMAP3, respectively.
In other words, the octopole axis is essentially unchanged whereas the quadrupole axis moves to the other side of the octopole axis, getting still closer
(from $13.0^\circ$ off to $11.5^\circ$), requiring a 1-in-50 fluke -- see \fig{axisofevil}. Moving the quadrupole axis in a random direction would typically {\it degrade} the alignment, so this is consistent with 
the hypothesis that the true alignment is still better and that this is partially masked by foregrounds.

Improved sensitivity from WMAP and Planck and also new cleaning techniques
will hopefully allow better removal and quantification of residual foreground contamination, which would reduce the main source of uncertainty in our measurements
and shed further light on the low-$\l$ CMB puzzles.

\end{document}